\documentclass{iaus}
\usepackage{graphicx}

\title[Lithium and its isotopic ratio in some roAp stars] 
{Lithium and its isotopic ratio $^6$Li/$^7$Li in the atmospheres of some
sharp-lined roAp stars}

\author[Shavrina et al.]   
{A. Shavrina$^1$, N. Polosukhina$^2$, S. Khan$^3$,  Ya. Pavlenko$^1$,
V. Khalack$^{1}$, G.A. Wade$^4$, P. Quinet$^5$, N. Mikhailitska$^1$,
A. Yushchenko$^{6,8}$, V. Gopka$^6$, A. Hatzes$^7$, D. Mkrtichian$^{6,8}$
\and A. Veles$^1$
}
\affiliation{
$^1$ Main Astronomical Observatory of NASU
 27 Zabolotnogo Str., 03680, Kyiv, Ukraine
\break email: shavrina@mao.kiev.ua, yp@mao.kiev.ua, khalack@mao.kiev.ua,
\break veles@mao.kiev.ua\\[\affilskip]
$^2$ Crimean Observatory, Ukraine
\break email: polo@crao.crimea.ua\\[\affilskip]
$^3$ Simpheropol University, Ukraine
\break email: serg@starsp.org\\[\affilskip]
$^4$ Physics Department, Royal Military College of Canada
\break email: gregg.wade@rmc.ca\\[\affilskip]
$^5$ Astrophysique et Spectroscopie, Universite de Mons-Hainaut, Belgium
\break email: Pascal.Quinet@umh.ac.be \\[\affilskip]
$^6$ Observatory of Odesa University, Ukraine
\break email: yua@odessa.net, gopka@arktur.tenet.odessa.ua \\[\affilskip]
$^7$ Th\"uringer Landersternwarte, Tautenburg, Germany
\break email: artie@tls-tautenburg.de\\[\affilskip]
$^8$ Astrophysical Research Center of the Structure and Evolution of 
the Cosmos (ARCSEC), Sejong University, Seoul 143-747, Korea
\break email: david@arcsec.sejong.ac.kr\\[\affilskip]
}

\pubyear{2004}
\volume{224}  
\pagerange{119--126}
\date{?? and in revised form ??}
\jname{The A-Star Puzzle}
\editors{J.\,Zverko, W.W.\,Weiss, J.\,\v{Z}i\v{z}\v{n}ovsk\'{y}, \& S.J.\,Adelman, eds.}
\begin{document}

\maketitle

\begin{abstract}
The lines of lithium at 6708~\AA\, and 6103~\AA\, are analyzed in high
resolution spectra of some sharp-lined and slowly rotating roAp stars. Three
spectral synthesis codes - STARSP, ZEEMAN2 and SYNTHM were used. New lines of
the rare earth elements from the DREAM database, and lines calculated on the
basis of the NIST energy levels were included. Magnetic splitting and other
line broadening processes were taken into account.  Enhanced  abundances
of lithium in the atmospheres of the stars studied are obtained for both the lithium
lines. High estimates of $^6$Li/$^7$Li ratio ($0.2\div0.5$) for the studied stars
can be explained by Galactic Cosmic Ray (GCR) production by to spallation
reactions and the preservation of the original $^6$Li and $^7$Li by the strong magnetic
fields.
\end{abstract}

\firstsection 
\section{Introduction}

In the framework of the project ``Lithium in CP stars", a significant series of
observations was obtained at ESO and CrAO (R=100000 and 50000 respectively,
1996--2001) for 5 rapidly oscillating Ap (roAp) stars: 33 Lib (HD\,137947),
$\gamma$ Equ (HD\,201601), HD\,134214, HD\,166473, HD\,101065, in the spectral
region 6680--6730~\AA. These series were supplemented by ESO (March 2004)
and SAO-BTA (April 2004) spectra with R=100000 and 60000. The observations show very
strong and non-variable resonance doublets of Li {\sc i} at 6708\AA. The spectra of 
these  roAp stars are group II in the classification of lithium roAp stars 
in accordance  with the Li {\sc i} line 6708\AA\, appearance over the phases
(\cite[Polosukhina, Kurtz, Hack, \etal\ 1999]{Polosukhina99})).

All these stars are characterized by sharp lines in their spectra, by the strong overabundance
of rare earth elements, and by magnetic fields from 2~kG up to 6.8~kG. The
sharp lines ($2\div3$~km~s$^{-1}$) in the spectra of these stars result from
small $v_e \sin{i}$. For the stars with short rotational periods the sharp lines
appear to be due to the combination of equatorial velocity $v_e$ and a significant
inclination angle $i$. For the stars with longer periods (of some years)
-- $\gamma$ Equ and 33 Lib -- the width of the lines is attributed by slow rotation.
(Note that the broadening of spectral lines due to rotation is not
distinguished from the broadening due to the rapid oscillations). Some of the
stars are therefore observed ``pole-on", and an observer always sees only one
hemisphere of these star. In this case the spectrum is essentially constant.

\begin{figure}[t]
\includegraphics[width=2.5in]{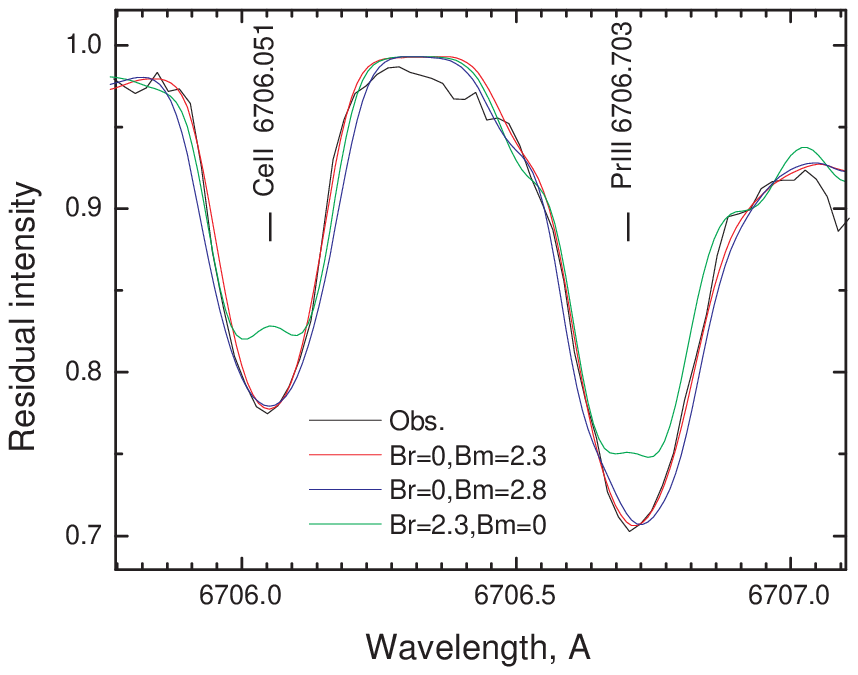}
\includegraphics[width=2.5in]{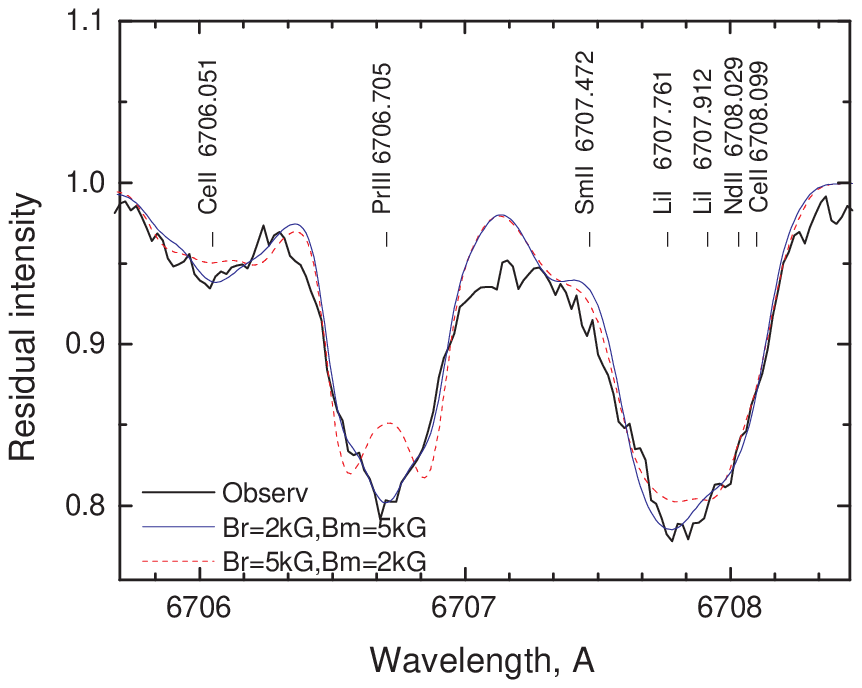}
\caption{Left: The estimation of field parameters from the lines Ce {\sc ii} 6706.051\AA\,
            and Pr {\sc iii} 6706.705\AA\ for HD\,101065, red solid line: $B_r=0$\,kG,
            $B_m=2.3$\,kG, $B_l=0$\,kG; blue solid line: $B_r=0$\,kG, $B_m=2.8$\,kG,
            $B_l=0$\,kG; green solid line: $B_r=2.3$\,kG, $B_m=0$\,kG, $B_l=0$\,kG;
         righ: for 33 Lib, blue line: $B_r=2$\,kG, $B_m=5$\,kG, $B_l=0$\,kG and
            red line: $B_r=5$\,kG, $B_m=2$\,kG, $B_l=0$\,kG.}
\label{fig1ab}
\end{figure}

\section{Synthetic spectra}

These stars with strong 6708\AA\, lithium doublets are very poorly studied. 
We study their spectra in detail in a narrow range near 6708\AA\, by
the method of synthetic spectra, taking into account Zeeman magnetic splitting
and blending by REE lines. The additional broadening, likely pulsational was 
described by the parameter $v \sin{i}$.

Spectral calculations for HD\,166473, $\gamma$\,Equ and 33~Lib were carried out
using the model atmospheres of \cite{Kurucz94} with parameters from the papers
of \cite{Gelbman00}, \cite{Ryabchikova97}, \cite{Ryabchikova99}. For HD\,101065
Pavlenko's model was used, as in the work of \cite{Shavrina03}. For synthetic
spectra calculations we applied the magnetic spectrum synthesis code SYNTHM
(\cite[Khan 2004]{Khan04}), which is similar to Piskunov code SYNTHMAG and was
tested in accordance with
the paper of \cite{Wade01}.  Also for initial calculations we used the
code STARSP of \cite{Tsymbal96} and in some cases the code ZEEMAN2 \cite{Wade01}.

\begin{figure}[thb]
   \includegraphics[width=2.5in]{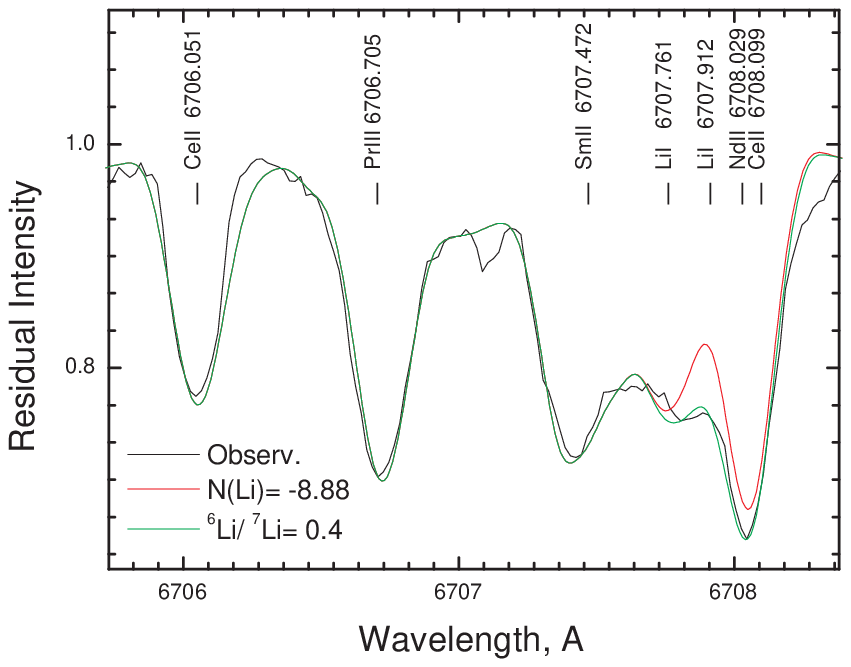} 
   \includegraphics[width=2.5in]{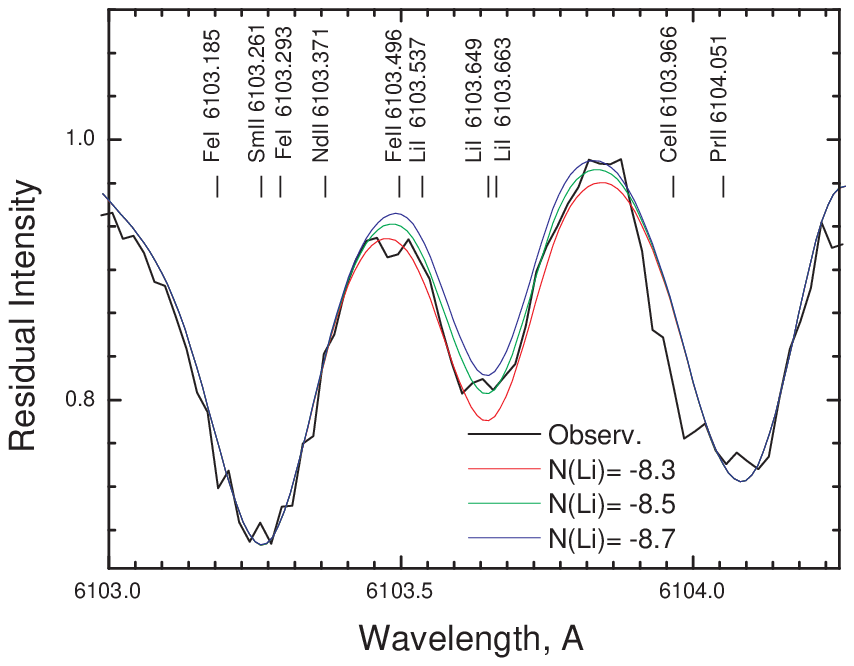}\\
\caption{Left: The fitting of observed and calculated
         spectra of HD\,101065 near 6708\,\AA: black: observed
         spectrum; red line: calculated spectrum taking
         into account lines of the main isotope $^{7}$Li only;
         green line: spectrum with the ratio $^{6}$Li/$^{7}$Li = 0.4. 
         The positions of those lines which are
         the main contributors in absorption are marked at
         the top of the figure;
         right: N(Li)=$-8.50 \pm 0.2$, $^{6}$Li/$^{7}$Li=0.4}
\label{fig2}
\end{figure}

    The simplified model of the magnetic field is characterized by radial(along
line of sight), meridional and longitudinal components of field $B_{\rm r},
B_{\rm m}, B_{\rm l}$
($B_{\rm l}$ = 0 always, as it is justified for the plane-parallel
model atmospheres), which were primarily determined from Fe {\sc ii} lines 6147\AA, 
6149\AA\, Ce {\sc ii} 6706.05 and Pr {\sc iii} 6706.70\AA\, (see Table~\ref{tab1}).

\begin{figure}[thb]
   \includegraphics[width=2.5in]{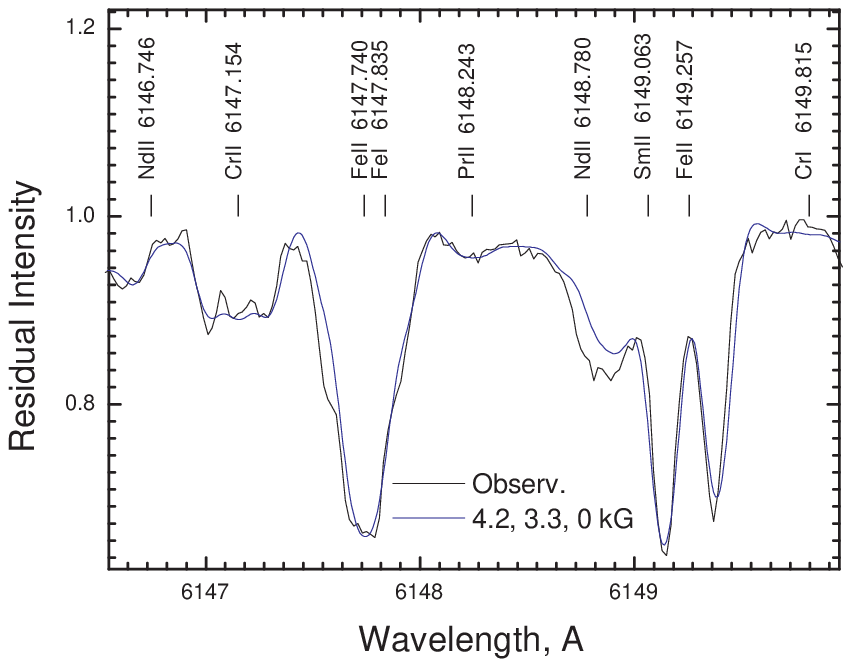}
   \includegraphics[width=2.5in]{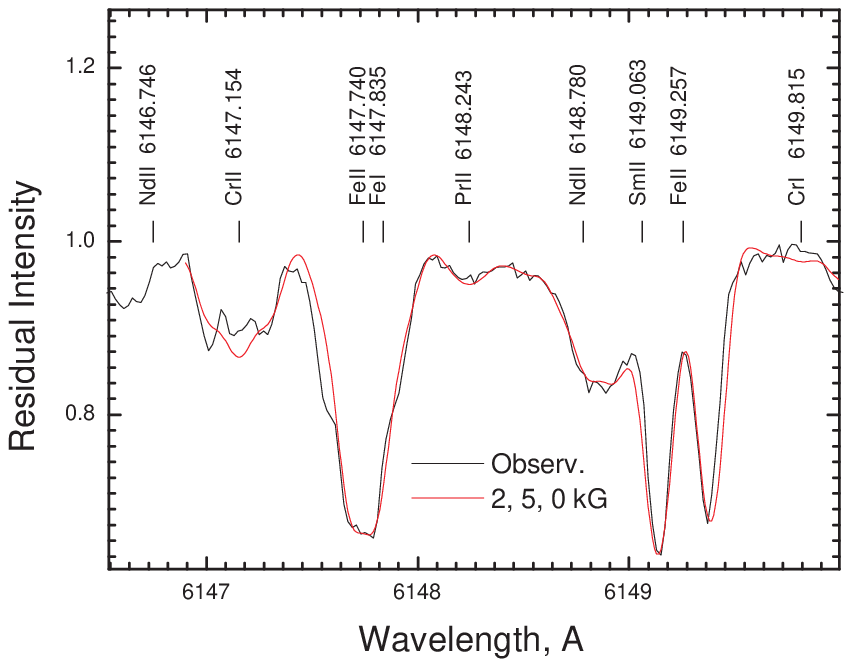}\\
\caption{Fe {\sc ii} lines 6147\,\AA\, and 6149\,\AA\, for 33\,Lib with
          magnetic field components. Left:~$B_r=4.2$\,kG, $B_m=3.3$\,kG, $B_l=0$\,kG;
         right: $B_r=2$\,kG, $B_m=5$\,kG and $B_l=0$\,kG like for Pr {\sc iii}
6708\,\AA\,
         (see Fig~\ref{fig1ab}b).}
\label{fig2a}
\end{figure}

\begin{table}\def~{\hphantom{0}}
  \begin{center}
  \caption{}
  \label{tab1}
  \begin{tabular}{lcccccc}\hline
   & HD\,101065 & HD\,134214 &  HD\,137949  &HD\,137949 & HD\,166473 & HD\,201601 \\\hline
T$_{eff}$/$\log{g}$/[m]& 6600/4.2/0& 7500/4.0/0& 7750/4.5/0&7250/4.5/0& 7750/4.0/0& 7750/4.0/0\\\hline
N(FeI) 6103\AA         &6.95 (Fe{\sc ii})&   7.60    &     8.00    & 7.80&    -     & 7.80\\
N(Fe{\sc ii}) 6149\AA  &   -       &   7.25    &     7.70    &7.80&    7.35   & 7.50\\
N(Li) 6708\AA        & 3.1       &   3.9     &     4.1     &3.6 &    3.7    & 3.8\\
N(Li) 6103\AA        & 3.5       &   4.1     &     4.4     &4.4 &     -     & 4.0\\
$^6$Li/$^7$Li 6708\AA& 0.4:      &   0.3:    &     0.2:    &0.3:&    0.5:   & 0.3:\\\hline
$B_r/ B_m/ B_l/$\ (kG)  & & & & & \\
Fe{\sc ii} 6149\AA         &   -       &-2.9/-1.7/0&  4.1/4.1/0  &4.2/3.3/0
&2.0/6.0/0 &3.5/2.6/0.8\\
Pr {\sc iii}I 6706.7\AA      & 0/2.3/0   &-2.3/-1.9/0&  2.0/5.0/0  &1.5/5.0/0 &
2.0/6.5/0 &2.7/3.5/0\\
CaI 6102.7\AA        & 0/2.4/0   &-1.7/-2.8/0&  3.0/4.0/0  &3.5/4.0/0  &     -    &
0/4.0/0\\
$v \sin i ({\rm km\,s^{-1}})$ & -   &      3.0 &      2.5    &2.5 &    3.0   &   2.5\\
 Fe {\sc ii}& -   &       &          & &      &   \\
$v \sin i ({\rm km\,s^{-1}})$ & 3.5&     2.0  &     4.0     &4.0 &    5.5   &   2.5\\
Pr {\sc iii}& &       &        & &       &  \\\hline
  \end{tabular}
 \end{center}
\end{table}

\section{REE lines with new atomic data}

We used the VALD (\cite[Kupka F., \etal\ 1999])) and DREAM 
\footnote{[\textit{http://www.umh.ac.be/$\tilde{\ }$astro/dream.shtml}]}
databases for atomic spectral lines. These data do not in fact
allow us to fit synthetic spectra to the observed ones for all stars studied. We
therefore calculated additional REE {\sc ii}-{\sc iii} lines using NIST energy levels and estimated
their "astrophysical" gf-values from the spectra of HD\,101065 using elemental
abundances from \cite{Cowley00}. As well, the theoretical gf-values for
important (under the lithium abundance determination) blending lines were
especially computed by P. Quinet with Cowan's code (\cite[Shavrina,
Polosukhina, Pavlenko, \etal\ 2003]{Shavrina03}).

\section{HD101065}

We present a new version of the spectraa analysis of the star HD\,101065 in the
lithium spectral ranges 6708\AA\, and 6103\AA\, using the new atomic data for
REE lines and the new magnetic synthesis code SYNTHM (\cite[Khan 2004]{Khan04}).
The lithium abundance estimates from 6708\AA\, and 6103\AA\, are 3.1~dex and
3.4~dex respectively, in the scale of $\log{N(H)}$=12.0~dex, and its isotopic
ratio $^6$Li/$^7$Li is about 0.4 (6708\AA) and 0.3 (6103\AA).

   \begin{figure}[t]
   \includegraphics[width=2.5in]{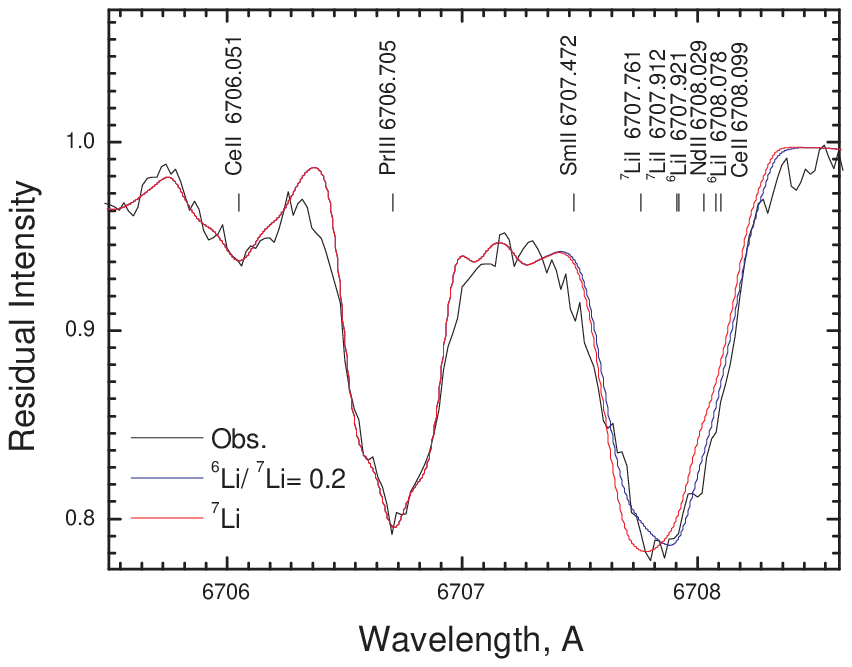}
   \includegraphics[width=2.5in]{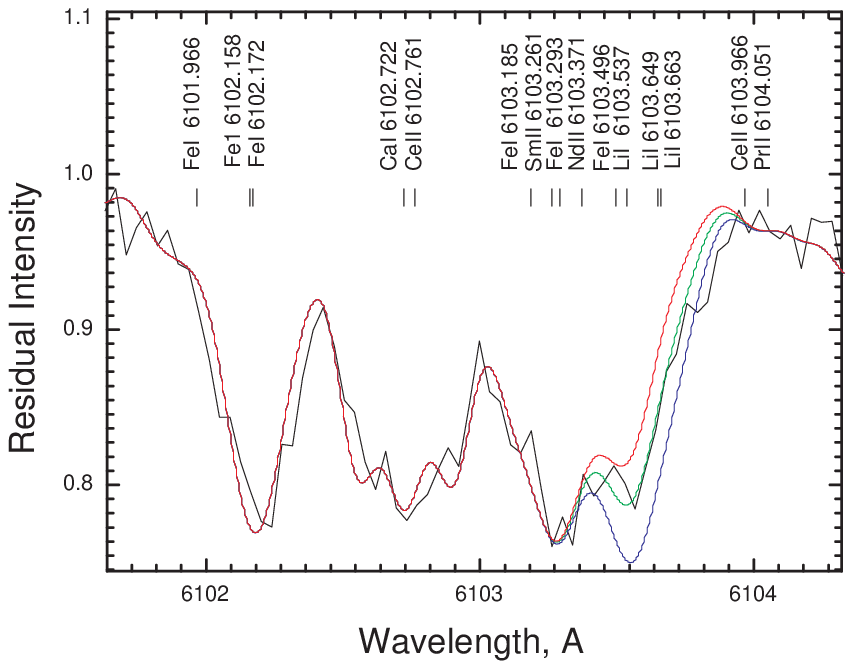}\\
   \caption{For 33~Lib a) Li {\sc i} 6708 \AA, 
        blue line: log N(Li) =-7.95, $^6$Li/$^7$Li =0.2;
        red line: log N(Li)= -7.88, only $^7$Li.
            b) Li {\sc i} 6103 \AA, green line: log N(Li)= -7.60 $\pm 0.3$,
    $^6$Li/$^7$Li =0.2 }
   \label{f1}
   \end{figure}

\section{Results}

Results of the work are presented in the Table. I the first line, the HD numbers
and in the second one - the parameters of used model atmospheres are given. 
The calculations for star HD 137949 were carried out for two model atmospheres 
in a possible effective temperature range - 7750/4.5/0 and 7250/4.5/0.
In six column for each star(model) we give the abundances of Fe I and Fe II
in the scale of log N(H)=12.0, derived from a group of the Fe I lines 
(6102-6103 \AA) and Fe II 6149 \AA (3,4 lines in the table). For HD 101065 
with weak Fe lines we use the abundance of Fe II from the paper of Cowley et al.
(2000) to take into account Fe II line 6103.496, which is near Li I lines 
(6103.538, 6103.649, 6103.664). 
The abundances of lithium determined from both 6708 A and 6103 A lines and 
isotopic ratio from 6708 A line are shown in 5-7 lines of table.
Under the solid line we give the parameters of magnetic field and vsini found 
from the fitting of Fe II 6149 \AA, Pr III 6706.7 and Ca I 6102.7 lines.
Last value of vsini was used for spectra calculations in both lithium lines 
ranges. Magnetic field parameters from Ca I 6102.7 were used in the 6103 \AA
range and ones from Pr III 6707.6 \AA - for 6708 \AA range.
 
\section{Conclusions}
\begin{itemize}

\item  The lithium abundance for all stars determined from the Li {\sc i}I 6103\AA\,
line is higher than the abundance determined from Li {\sc i}I 6708\AA. This may be 
evidence of vertical lithium stratification, an abnormal temperature
distribution, or consistent unidentified blending with the 6103\AA\ line.

\item Our work on two roAp stars, HD\,83368 and HD\,60435 (\cite[Shavrina,
Polosukhina, Zverko, \etal\ 2001]{Shavrina01}) provides evidence of an enhanced
lithium abundance near the magnetic field poles. We can expect similar
effects in sharp-lined roAp stars. The high lithium abundance for
all stars determined from the Li {\sc i}I lines and the estimates of $^6$li/$^7$Li
ratio ($0.2\div0.5$) can be explained by the Galactic Cosmic Ray (GCR)
production due to spallation reactions with ISM in the areas of these stars 
formation and preservation of original both $^6$Li and
$^7$Li by the strong magnetic fields of these stars. The values of the $^6$Li/$^7$Li ratio
expected from GCR production are about $0.5\div0.8$ (\cite[Knauth et al.
2003]{Knauth03}, \cite[Webber, Lukasiak, McDonald, 2002]{Webber02}).

\item The new laboratory and theoretical gf-values for REE lines are necessary
in order to refine our estimates of lithium abundances and the isotopic ratio.

\end{itemize}

\begin{acknowledgments}
The authors are grateful to Dr. J. Zverko and Dr. J. \v Zi\v z\v novsk\'y for
their useful comments.
A. Shavrina, N. Polosukhina, V. Khalack and V. Gopka
would like also to express  gratitude to the Local
Organizing Committee and IAU for financial support.
\end{acknowledgments}

\end{document}